
\input harvmac

\Title{\vbox{\baselineskip12pt
\hbox{McGill/95-24}
\hbox{CERN-TH/95-156}
\hbox{hep-th/9506086}
}}
{\vbox{\centerline {Spacetime Supersymmetry and Duality
in String Theory}}}
\footnote{}{Talk given at MRST'95 Meeting,
May 8-9, 1995,
Rochester University}
\centerline{Ramzi R.~Khuri}
\bigskip\centerline{{\it Physics Department, McGill University}}
\centerline{\it Montreal, PQ, H3A 2T8 Canada}
\bigskip
\centerline{and}
\bigskip\centerline{{\it CERN,
 CH-1211, Geneva 23, Switzerland}}
\vskip .3in
I discuss the role of spacetime supersymmetry in the interplay
between strong/weak coupling duality and target space duality
in
string theory which arises in string/string duality.
This can be seen via the construction
of string soliton solutions which in $N=4$ compactifications
of heterotic string theory break more than
$1/2$ of the spacetime supersymmetries but whose analogs in
$N=2$ and $N=1$
compactifications break precisely $1/2$ of the spacetime
supersymmetries.
As a result, these solutions may be interpreted as stable
solitons
in the latter two cases, and correspond to
Bogomol'nyi-saturated states
in their respective spectra.

\vskip .3in
\Date{\vbox{\baselineskip12pt
\hbox{McGill/95-24}
\hbox{CERN-TH/95-156}
\hbox{June 1995}}}

\def\sqr#1#2{{\vbox{\hrule height.#2pt\hbox{\vrule width
.#2pt height#1pt \kern#1pt\vrule width.#2pt}\hrule height.#2pt}}}

\lref\prep{M. J. Duff, R. R. Khuri and J. X. Lu,
 NI-94-017, CTP/TAMU-67/92,
McGill/94/53, CERN-TH.7542/94, hepth/9412184
(to appear in Physics Reports).}

\lref\fabjr{M. Fabbrichesi, R. Jengo and K. Roland, Nucl.
Phys. {\bf B402} (1993) 360.}

\lref\jxthesis{J. X. Lu, {\it Supersymmetric Extended
Objects}, Ph.D. Thesis, Texas A\&M University (1992),
UMI {\bf 53 08B}, Feb. 1993.}

\lref\coleman{S. Coleman, ``Classical Lumps and Their Quantum
Descendants'', in {\it New Phenomena in Subnuclear Physics},
ed A. Zichichi (Plenum, New York, 1976).}

\lref\jackiw{R. Jackiw, Rev. Mod. Phys. {\bf 49} (1977) 681.}

\lref\hink{M. B. Hindmarsh and T. W. B. Kibble, SUSX-TP-94-74,
IMPERIAL/TP/94-95/5, NI 94025, hepph/9411342.}

\lref\senmod{A. Sen, Mod. Phys. Lett. {\bf A8} (1993) 2023.}

\lref\kikyam{K. Kikkawa and M. Yamasaki, Phys. Lett. {\bf B149}
(1984) 357.}

\lref\sakai{N. Sakai and I. Senda, Prog. Theor. Phys. {\bf 75}
(1986) 692.}

\lref\bush{T. Busher, Phys. Lett. {\bf B159} (1985) 127.}

\lref\nair{V. Nair, A. Shapere, A. Strominger and F. Wilczek,
Nucl. Phys. {\bf B322} (1989) 167.}

\lref\duff{M. J. Duff, Nucl. Phys. {\bf B335} (1990)
610.}

\lref\tseyv{A. A. Tseytlin and C. Vafa, Nucl. Phys. {\bf B372}
(1992) 443,}

\lref\tseycqg{A. A. Tseytlin, Class. Quantum Grav. {\bf 9}
(1992) 979.}

\lref\givpr{A. Giveon, M. Porrati and E. Rabinovici,
(to appear in Phys. Rep. C).}

\lref\gibkal{G. W. Gibbons and R. Kallosh, NI-94003,
hepth/9407118.}

\lref\vafw{C. Vafa and E. Witten, Nucl. Phys. {\bf B431}
(1994) 3.}

\lref\seiw{N. Seiberg and E. Witten, Nucl. Phys. {\bf B426}
(1994) 19.}

\lref\seiwone{N. Seiberg and E. Witten, Nucl. Phys. {\bf B431}
(1994) 484.}

\lref\cerdfv{A. Ceresole, R. D'Auria, S. Ferrara and
A. Van Proeyen, CERN-TH 7510/94, POLFIS-TH.08/94,
UCLA 94/TEP/45, KUL-TF-94/44, hepth/9412200.}

\lref\cerdfvone{A. Ceresole, R. D'Auria, S. Ferrara and
A. Van Proeyen, CERN-TH 7547/94, POLFIS-TH.01/95,
UCLA 94/TEP/45, KUL-TF-95/4, hepth/9502072.}

\lref\gauh{J. Gauntlett and J. H. Harvey, EFI-94-30,
hepth/9407111.}

\lref\frak{P. H. Frampton and T. W. Kephart, IFP-708-UNC,
VAND-TH-94-15.}

\lref\hawhr{S. W. Hawking, G. T. Horowitz and S. F. Ross,
NI-94-012, DAMTP/R 94-26, UCSBTH-94-25, gr-qc/9409013.}

\lref\ellmn{J. Ellis, N. E. Mavromatos and D. V. Nanopoulos,
Phys. Lett. {\bf B278} (1992) 246.}

\lref\kaln{S. Kalara and N. Nanopoulos, Phys. Lett. {\bf B267}
(1992) 343.}

\lref\glasgow{M. J. Duff, NI-94-016, CTP-TAMU-48/94,
hepth/9410210.}

\lref\dufnew{M. J. Duff, NI-94-033, CTP-TAMU-49/94,
hepth/9501030.}

\lref\bhs{R. R. Khuri, Helv. Phys. Acta {\bf 67} (1994) 884.}

\lref\cecfg{S. Cecotti, S. Ferrara and L. Girardello,
Int. J. Mod. Phys. {\bf A4} (1989) 2475.}

\lref\ferquat{S. Ferrara and S. Sabharwal,
Nucl. Phys. {\bf B332} (1990) 317.}

\lref\fere{R. C. Ferrell and D. M. Eardley,
Phys. Rev. Lett. {\bf 59} (1987) 1617.}

\lref\reyt{S. J. Rey and T. R. Taylor, Phys. Rev. Lett. {\bf 71}
(1993) 1132.}

\lref\senzwione{A. Sen and B. Zwiebach, Nucl. Phys.
{\bf B414} (1994) 649.}

\lref\senzwitwo{A. Sen and B. Zwiebach, Nucl. Phys.
{\bf B423} (1994) 580.}

\lref\bddo{T. Banks, A. Dabholkar, M. R. Douglas and
M. O'Loughlin, Phys. Rev. {\bf D45} (1992) 3607.}

\lref\bos{T. Banks, M. O'Loughlin and A. Strominger,
Phys. Rev. {\bf D47} (1993) 4476.}

\lref\kir{E. Kiritsis, Nucl. Phys. {\bf B405} (1993) 109.}

\lref\dufr{M. J. Duff and J. Rahmfeld, Phys. Lett. {\bf B345} (1995)
441.}

\lref\hult{C. M. Hull and P. K. Townsend, Nucl. Phys. {\bf B438}
(1995) 109.}

\lref\duffkk{M. J. Duff, NI-94-015, CTP-TAMU-22/94,
hepth/9410046.}

\lref\bakone{I. Bakas, Nucl. Phys. {\bf B428} (1994) 374.}

\lref\baktwo{I. Bakas, Phys. Lett. {\bf B343} (1995) 103.}

\lref\baksfet{I. Bakas and K. Sfetsos, CERN-TH-95-16,
hepth/9502065.}

\lref\back{C. Bachas and E. Kiritsis, Phys. Lett. {\bf B325}
(1994) 103.}

\lref\bko{E. Bergshoeff, R. Kallosh and T. Ortin,
UG-8/94, SU-ITP-94-19, QMW-PH-94-13, hepth/9410230.}

\lref\vilsh{A. Vilenkin and E. P. Shellard,
{\it Cosmic String and Other Topological Defects},
(Cambridge University Press, 1994).}

\lref\bfrm{M. Bianchi, F. Fucito, G. C. Rossi and M. Martellini,
hepth/9409037.}

\lref\ght{G. W. Gibbons, G. T. Horowitz and P. K. Townsend,
R/94/28, UCSBTH-94-35, hepth/9410073.}

\lref\dufkmr{M. J. Duff, R. R. Khuri, R. Minasian and
J. Rahmfeld, Nucl. Phys. {\bf B418} (1994) 195.}

\lref\sentd{A. Sen, Nucl. Phys. {\bf B434} (1995) 179.}

\lref\maha{J. Maharana, NI-94023, hepth/9412235.}

\lref\gresss{M. B. Green and J. Schwarz, Phys. Lett. {\bf B136}
(1984) 367.}

\lref\sie{W. Siegel, Phys. Lett. {\bf B128} (1983) 397.}

\lref\dir {P. A. M. Dirac, Pro. R. Soc. {\bf A133} (1931) 60.}

\lref\tho {G. t'Hooft, Nucl. Phys. {\bf B79} (1974) 276.}

\lref\pol {A. M. Polyakov, Sov. Phys. JETP Lett. {\bf 20}
(1974) 194.}

\lref\mono {C. Montonen and D. Olive, Phys. Lett. {\bf B72}
(1977) 117.}

\lref\col {S. Coleman, Phys. Rev. {\bf D11} (1975) 2088.}

\lref\grohmr {D. J. Gross, J. A. Harvey, E. Martinec and
 R. Rohm,
Nucl. Phys. {\bf B256} (1985) 253.}

\lref\ginone {P. Ginsparg,
 {\it Conformal Field Theory}, Lectures given
at Trieste Summer School, Trieste, Italy, 1991.}

\lref\calhstwo {C. Callan, J. Harvey and A. Strominger,
 {\it Supersymmetric String Solitons}, Lectures given
at Trieste Summer School, Trieste, Italy, 1991.}

\lref\sch {J. H. Schwarz, {\it Supersymmetry and Its
 Applications}
ed G. W. Gibbons {\it et al} (Cambridge University Press,
1986).}

\lref\huglp {J. Hughes, J. Liu and J. Polchinski, Phys. Lett.
{\bf B180} (1986)
370.}

\lref\berst {E. Bergshoeff, E. Sezgin and P. K. Townsend,
 Phys. Lett. {\bf B189} (1987) 75.}

\lref\achetw {A. Achucarro, J. Evans, P. K.  Townsend and
D. Wiltshire,
 Phys.
Lett. {\bf B198} (1987) 441.}

\lref\belpst {A. A. Belavin, A. M. Polyakov, A. S. Schwartz and
Yu. S. Tyupkin,
Phys. Lett. {\bf B59} (1975) 85.}

\lref\oset {D. O'Se and D. H. Tchrakian, Lett. Math. Phys.
{\bf 13} (1987) 211.}

\lref\groks {B. Grossman, T. W. Kephart and J. D. Stasheff,
 Commun. Math. Phys. {\bf 96} (1984) 431;
Commun. Math. Phys. {\bf 100} (1985) 311.}

\lref\grokstwo {B. Grossman, T. W. Kephart and J. D. Stasheff,
 Phys. Lett. {\bf B220}
 (1989)  431.}

\lref\tch{D. H. Tchrakian, Phys. Lett. {\bf B150}  (1985)  360.}

\lref\fubn{S. Fubini and H. Nicolai,
 Phys. Lett. {\bf B155}  (1985)  369}

\lref\fain{D. B. Fairlie and J. Nuyts,
 J. Phys. {\bf A17}  (1984)  2867.}

\lref\str {A. Strominger,  Nucl. Phys. {\bf B343}
(1990) 167.}

\lref\duflhs {M. J. Duff and J. X. Lu,
Phys. Rev. Lett. {\bf 66} (1991) 1402.}

\lref\godo {P. Goddard and D. Olive, Rep. Prog. Phys. {\bf 41}
(1978) 1357.}

\lref\colone {S. Coleman, Proc. 1975 Int. School on Subnuclear
Physics,
Erice, ed A. Zichichi  (Plenum, New York, 1977); Proc. 1981
Int. School
on Subnuclear Physics, Erice, ed A. Zichichi (Plenum, New York,
1983).}

\lref\wuy {T. T. Wu and C. N. Yang, Nucl. Phys. {\bf B107}
(1976) 365.}

\lref\dufldl {M. J. Duff and J. X. Lu, Class. Quantum Grav.
{\bf 9} (1992) 1.}

\lref\tei {C. Teitelboim,
Phys. Lett. {\bf B167} (1986) 69.}

\lref\nep {R. I. Nepomechie,
Phys. Rev. {\bf D31} (1984) 1921.}

\lref\raj {R. Rajaraman, {\it Solitons and Instantons}
(North--Holland, Amsterdam, 1982).}

\lref\wito {E. Witten and D. Olive, Phys. Lett.
{\bf B78} (1978) 97.}

\lref\pras {M. K. Prasad and C. M. Sommerfield, Phys. Rev. Lett.
{\bf 35} (1975) 760.}

\lref\corg {E. Corrigan and P. Goddard, Commun. Math. Phys.
{\bf 80} (1981)
575.}

\lref\godno {P. Goddard, J. Nuyts and D. Olive, Nucl. Phys.
{\bf B125} (1977)
1.}

\lref\osb {H. Osborn, Phys. Lett. {\bf B83} (1979) 321.}

\lref\egugh {T. Eguchi, P. B. Gilkey and A. J. Hanson,
Phys. Rep. {\bf 66}
(1980) 213.}

\lref\corf {E. F. Corrigan and D. B. Fairlie, Phys. Lett.
{\bf B67} (1977) 69.
}

\lref\atidhm {M. F. Atiyah, V. G. Drinfeld, N. J. Hitchin and
Y. I. Manin,
Phys. Lett. {\bf A65} (1978) 185.}

\lref\dun {A. R. Dundarer,
 Mod. Phys. Lett. {\bf A5} (1991) 409.}

\lref\cha {A. H. Chamseddine, Phys. Rev. {\bf D24} (1981) 3065.}

\lref\berrwv {E. A. Bergshoeff, M. de Roo, B. de Wit and
 P. van Nieuwenhuizen,
Nucl. Phys. {\bf B195}  (1982)  97}

\lref\cham{G. F. Chapline and N. S. Manton, Phys. Lett.
{\bf B120} (1983) 105.}

\lref\gatn {S. J. Gates and H. Nishino, Phys. Lett. {\bf B173}
(1986) 52.}

\lref\sala{A. Salam and E. Sezgin, Physica Scripta {\bf 32}
(1985) 283.}

\lref\duf {M. J. Duff,  Class.
 Quantum  Grav. {\bf 5} (1988) 189.}

\lref\duflfb {M. J. Duff and J. X. Lu, Nucl. Phys. {\bf B354}
(1991) 141.}

\lref\dabghr {A. Dabholkar, G. W. Gibbons, J. A. Harvey and
F. Ruiz Ruiz,
 Nucl. Phys. {\bf B340} (1990) 33.}

\lref\berdps {E. Bergshoeff, M. J. Duff, C. N. Pope and
E. Sezgin,
 Phys. Lett.
{\bf B199} (1987) 69.}

\lref\tow {P. K. Townsend,
Phys. Lett. {\bf B202} (1988) 53.}

\lref\dufhis {M. J. Duff, P. S. Howe, T. Inami and K. Stelle,
 Phys. Lett.
{\bf B191} (1987) 70.}

\lref\dufs {M. J. Duff and K. Stelle, Phys. Lett. {\bf B253}
(1991) 113.}

\lref\berst {E. Bergshoeff, E. Sezgin and P. K. Townsend, Ann.
 Phys. {\bf 199}
(1990) 340.}

\lref\duflrsfd {M. J. Duff and J. X. Lu, Nucl. Phys. {\bf B354}
(1991) 129.}

\lref\gresone {M. Green and J. Schwarz, Phys. Lett. {\bf B151}
(1985) 21.}

\lref\bercgw {C. W. Bernard, N. H. Christ, A. H. Guth and
E. J. Weinberg,
 Phys. Rev.
 {\bf D16} (1977) 2967.}

\lref\hars {J. Harvey and A. Strominger,
 Phys. Rev. Lett. {\bf 66} (1991) 549.}

\lref\gres {M. Green and J. Schwarz, Phys. Lett. {\bf B149}
(1984) 117.}

\lref\elljm {J. Ellis, P. Jetzer and L. Mizrachi,
Nucl. Phys. {\bf B303} (1988)
1.}

\lref\dixds {J. Dixon, M. J. Duff and E. Sezgin, Phys. Lett.
{\bf B279} (1992) 265.}

\lref\berrs {E. Bergsheoff, M. Rakowski and E. Sezgin, Phys.
 Lett. {\bf
B185}  (1987)  371}

\lref\berd{E. Bergsheoff and M. de Roo, Nucl. Phys. {\bf B328}
(1989)
439}

\lref\dersw{M. de Roo, H. Suelmann and A. Wiedemann, preprint
UG--1/92  (1992).}

\lref\gresw {M. Green, J. Schwarz and E. Witten,
{\it Superstring
theory} (Cambridge University Press, 1987).}

\lref\duflloop {M. J. Duff and J. X. Lu, Nucl. Phys. {\bf B357}
  (1991)  534.}

\lref\ven {G. Veneziano,
Europhys. Lett. {\bf 2}  (1986)  199.}

\lref\cain {Y. Cai and C. A. Nunez, Nucl. Phys. {\bf B287}
(1987)  41}

\lref\gros{D. J. Gross and J. Sloan, Nucl. Phys. {\bf B291}
(1987)  41.}

\lref\ellm {J. Ellis and L. Mizrachi,
  Nucl. Phys. {\bf B327}  (1989)  595.}

\lref\calfmp {C. G. Callan, D. Friedan, E. J. Martinec and
M. J. Perry,
Nucl. Phys. {\bf B262}  (1985)  593.}

\lref\grestwo {M. Green and J. Schwarz, Phys. Lett. {\bf B173}
 (1986)  52.}

\lref\lin {U. Lindstrom, in Supermembranes and Physics in 2 + 1
Dimensions, ed. M. J. Duff, C. N. Pope and E. Sezgin
(World Scientific,
Singapore) (1990).}

\lref\dufone {M. J. Duff, Class. Quantum Grav. {\bf 6}  (1989)
1577.}

\lref\callny {C. Callan, C. Lovelace, C. Nappi and S. Yost,
Nucl. Phys.
{\bf B308}  (1988)  221.}

\lref\frat {E. Fradkin and A. Tseytlin, Phys. Lett. {\bf B158}
(1985)  316.}

\lref\duflselft {M. J. Duff and J. X. Lu, Phys. Lett.
{\bf B273}  (1991)  409.}

\lref\hors {G. Horowitz and A. Strominger, Nucl. Phys.
{\bf B360}  (1991) 197. }

\lref\witone {E. Witten,  Phys. Lett. {\bf B86}  (1979)
283.}

\lref\schone {J. Schwarz, Nucl. Phys. {\bf B226}  (1983)  269.}

\lref\zwa {D. Zwanziger, Phys. Rev. {\bf 176}  (1968)  1480,
1489.}

\lref\schwing {J. Schwinger, Phys. Rev. {\bf 144} (1966) 1087;
{\bf 173}
(1968) 1536.}

\lref\gibt{G.W. Gibbons and P.K. Townsend, Phys. Rev. Lett.
{\bf 71} (1993)
3754.}

\lref\duflblacks {M. J. Duff and J. X. Lu,
 Nucl. Phys. {\bf B416} (1994) 301.}

\lref\dufklsin {M. J. Duff, R. R. Khuri and J. X. Lu,
 Nucl. Phys. {\bf B377}
(1992) 281.}

\lref\calk {C.~G.~Callan and R.~ R.~Khuri,
Phys. Lett. {\bf B261} (1991) 363.}

\lref\gib {G. W. Gibbons, Nucl. Phys. {\bf B207} (1982) 337.}

\lref\gibm {G. W. Gibbons and K. Maeda, Nucl. Phys. {\bf B298}
(1988) 741.}

\lref\rey{S. J. Rey, in Proceedings of Tuscaloosa
Workshop on Particle Physics, (Tuscaloosa, Alabama, 1989).}

\lref\reyone {S. J.~Rey, Phys. Rev. {\bf D43} (1991) 526.}

\lref\antben {I.~Antoniadis, C.~Bachas, J.~Ellis and
 D.~V.~Nanopoulos,
Phys. Lett. {\bf B211} (1988) 393.}

\lref\antbenone {I.~Antoniadis, C.~Bachas, J.~Ellis and
D.~V.~Nanopoulos,
Nucl. Phys. {\bf B328} (1989) 117.}

\lref\mett {R.~R.~Metsaev and A.~A.~Tseytlin, Phys. Lett.
{\bf B191} (1987) 354.}

\lref\mettone {R.~R.~Metsaev and A.~A.~Tseytlin,
Nucl. Phys. {\bf B293} (1987) 385.}

\lref\calkp {C.~G.~Callan,
I.~R.~Klebanov and M.~J.~Perry, Nucl. Phys. {\bf B278} (1986)
78.}

\lref\lov {C.~Lovelace, Phys. Lett. {\bf B135} (1984) 75.}

\lref\friv {B.~E.~Fridling and A.~E.~M.~Van de Ven,
Nucl. Phys. {\bf B268} (1986) 719.}

\lref\gepw {D.~Gepner and E.~Witten, Nucl. Phys. {\bf B278}
(1986) 493.}

\lref\din {M.~Dine, Lectures delivered at
TASI 1988, Brown University (1988) 653.}

\lref\berdone {E.~A.~Bergshoeff and M.~de Roo, Phys. Lett.
{\bf B218} (1989)
210.}

\lref\calhs{C.~G.~Callan, J.~A.~Harvey and A.~Strominger,
Nucl. Phys.
{\bf B359} (1991) 611.}

\lref\calhsone{C.~G.~Callan, J.~A.~Harvey and A.~Strominger,
Nucl. Phys.
{\bf B367} (1991) 60.}

\lref\thoone{G.~'t~Hooft, Phys. Rev. Lett. {\bf 37} (1976) 8.}

\lref\wil{F.~Wilczek, in
{\it Quark confinement and field theory},
Eds. D.~Stump and D.~Weingarten, (John Wiley and Sons, New York,
1977).}

\lref\jacnr{R.~Jackiw, C.~Nohl and C.~Rebbi, Phys. Rev.
{\bf D15} (1977)
1642.}

\lref\khuinst{R.~R.~Khuri, Phys. Lett.
{\bf B259} (1991) 261.}

\lref\khumant{R.~R.~Khuri, Nucl. Phys.
 {\bf B376} (1992) 350.}

\lref\khumono{R.~R.~Khuri,
 Phys. Lett. {\bf B294} (1992) 325.}

\lref\khumonscat{R.~R.~Khuri,
Phys. Lett. {\bf B294} (1992) 331.}

\lref\khumonex{R.~R.~Khuri,
Nucl. Phys. {\bf B387} (1992) 315.}

\lref\khumonin{R.~R.~Khuri,
 Phys. Rev. {\bf D46} (1992) 4526.}

\lref\khugeo{R.~R.~Khuri,
Phys. Lett. {\bf 307} (1993) 302.}

\lref\khuscat{R.~R.~Khuri,
 Nucl. Phys. {\bf B403} (1993) 335.}

\lref\khuwind {R.~R.~Khuri, Phys. Rev. {\bf D48} (1993) 2823.}

\lref\gin{P.~Ginsparg, Lectures delivered at
Les Houches summer session, June 28--August 5, 1988.}

\lref\alljj{R. W. Allen, I. Jack and D. R. T. Jones,
 Z. Phys. {\bf C41}
(1988) 323.}

\lref\sev{A. Sevrin, W. Troost and A. van Proeyen,
Phys. Lett. {\bf B208} (1988) 447.}

\lref\schout{K. Schoutens, Nucl. Phys. {\bf B295} [FS21] (1988)
634.}

\lref\harl{J.~A.~Harvey and J.~Liu, Phys. Lett. {\bf B268}
(1991) 40.}

\lref\man{N.~S.~Manton, Nucl. Phys. {\bf B126} (1977) 525.}

\lref\manone{N.~S.~Manton, Phys. Lett. {\bf B110} (1982) 54.}

\lref\mantwo{N.~S.~Manton, Phys. Lett. {\bf B154} (1985) 397.}

\lref\atihone{M.~F.~Atiyah and N.~J.~Hitchin, Phys. Lett.
{\bf A107}
(1985) 21.}

\lref\atihtwo{M.~F.~Atiyah and N.~J.~Hitchin, {\it The Geometry
and
Dynamics of Magnetic Monopoles}, (Princeton University Press,
1988).}

\lref\polc{J.~Polchinski, Phys. Lett. {\bf B209} (1988) 252.}

\lref\gibhp{G.~W.~Gibbons and S.~W.~Hawking, Phys. Rev.
{\bf D15}
(1977) 2752.}

\lref\gibhpone{G.~W.~Gibbons, S.~W.~Hawking and M.~J.~Perry,
 Nucl. Phys.
{\bf B318} (1978) 141.}

\lref\brih{D.~Brill and G.~T.~Horowitz, Phys. Lett. {\bf B262}
(1991)
437.}

\lref\gids{S.~B.~Giddings and A.~Strominger, Nucl. Phys.
{\bf B306}
(1988) 890.}

\lref\gidsone{S.~B.~Giddings and A.~Strominger, Phys. Lett.
{\bf B230}
(1989) 46.}

\lref\canhsw{P.~Candelas, G.~T.~Horowitz, A.~Strominger and
E.~Witten,
Nucl. Phys. {\bf B258} (1984) 46.}

\lref\bog{E.~B.~Bogomolnyi, Sov. J. Nucl. Phys. {\bf 24} (1976)
449.}

\lref\war{R.~S.~Ward, Comm. Math. Phys. {\bf 79} (1981) 317.}

\lref\warone{R.~S.~Ward, Comm. Math. Phys. {\bf 80} (1981) 563.}

\lref\wartwo{R.~S.~Ward, Phys. Lett. {\bf B158} (1985) 424.}

\lref\grop{D.~J.~Gross and M.~J.~Perry, Nucl. Phys. {\bf B226}
(1983)
29.}

\lref\ash{{\it New Perspectives in Canonical Gravity}, ed.
A.~Ashtekar,
(Bibliopolis, 1988).}

\lref\lic{A.~Lichnerowicz, {\it Th\' eories Relativistes de la
Gravitation et de l'Electro-magnetisme}, (Masson, Paris 1955).}

\lref\gol{H.~Goldstein, {\it Classical Mechanics},
Addison-Wesley,
1981.}

\lref\ros{P.~Rossi, Physics Reports, 86(6) 317-362.}

\lref\dixdp{J.~A.~Dixon, M.~J.~Duff and J.~C.~Plefka,
 Phys. Rev.
Lett.
{\bf 69} (1992) 3009.}

\lref\chad{J.~M.~Charap and M.~J.~Duff, Phys. Lett. {\bf B69}
(1977) 445.}

\lref\dufkexst{M.~J.~Duff and R.~R.~Khuri,
Nucl. Phys. {\bf B411} (1994) 473.}

\lref\khubifb{R.~R.~Khuri,
Phys. Rev. {\bf D48} (1993) 2947.}

\lref\khustab{R.~R.~Khuri, Phys. Lett. {\bf B307} (1993) 298.}

\lref\sor{R.~D.~Sorkin, Phys. Rev. Lett. {\bf 51} (1983) 87.}

\lref\dabh{A.~Dabholkar and J.~A.~Harvey,
 Phys. Rev. Lett. {\bf 63} (1989) 478.}

\lref\fels{A.~G.~Felce and T.~M.~Samols, Phys. Lett.
{\bf B308} (1993) 30.}

\lref\dufipss{M.~J.~Duff, T.~Inami, C.~N.~Pope, E.~Sezgin and
K.~S.~Stelle,
Nucl. Phys. {\bf B297}
(1988) 515.}

\lref\fujku{K.~Fujikawa and J.~Kubo,
 Nucl. Phys. {\bf B356} (1991) 208.}

\lref\cvet{M.~Cveti\v c, Phys. Rev. Lett. {\bf 71} (1993) 815.}

\lref\cvegs{M.~Cveti\v c, S. Griffies and H. H. Soleng, Phys.
 Rev. Lett. {\bf 71} (1993) 670; Phys. Rev. {\bf D48} (1993)
 2613.}

\lref\la{H. S. La, Phys. Lett. {\bf B315} (1993) 51.}

\lref\gresvy{B.~R.~Greene, A.~Shapere, C.~Vafa and S.~T.~Yau,
 Nucl. Phys.
{\bf B337} (1990) 1.}

\lref\fonilq{A.~Font, L.~Ib\'a\~nez, D.~Lust and F.~Quevedo,
Phys. Lett.
{\bf B249} (1990) 35.}

\lref\bin{P.~Bin\'etruy, Phys. Lett. {\bf B315} (1993) 80.}

\lref\koun{C.~Kounnas, in {\it Proceedings of INFN Eloisatron
Project, 26th Workshop: ``From Superstrings to Supergravity",
 Erice, Italy,
Dec. 5-12, 1992}, Eds. M.~Duff, S.~Ferrara and R.~Khuri,
(World Scientific, 1994).}

\lref\duftv{M.J. Duff, P.K. Townsend and P. van Nieuwenhuizen,
Phys. Lett. {\bf B122} (1983) 232.}

\lref\dufgt{M.J. Duff, G.W. Gibbons and P.K. Townsend
Phys. Lett. {\bf B} (1994) }

\lref\guv{R. G\"uven, Phys. Lett. {\bf B276} (1992) 49.}

\lref\guven{R. G\"uven, Phys. Lett. {\bf B212} (1988) 277.}

\lref\dobm{P.~Dobiasch and D.~Maison, Gen. Rel. Grav.
{\bf 14} (1982) 231.}

\lref\chod{A.~Chodos and S.~Detweiler, Gen. Rel. Grav.
{\bf 14} (1982) 879.}

\lref\pol{D.~Pollard, J. Phys. {\bf A16} (1983) 565.}

\lref\duffkr{M.~J.~Duff, S.~Ferrara, R.~R.~Khuri and J.~Rahmfeld,
 hep-th/9506057.}

\lref\lu{J. X. Lu,
 Phys. Lett. {\bf B313} (1993) 29.}

\lref\grel{R. Gregory and R. Laflamme, Phys. Rev. Lett.
{\bf 70} (1993) 2837.}

\lref\rom{L. Romans, Nucl. Phys. {\bf B276} (1986) 71.}

\lref\sala {A. Salam and E. Sezgin, {\it Supergravities in
Diverse Dimensions},
(North Holland/World Scientific, 1989).}

\lref\strath {J. Strathdee, Int. J. Mod. Phys. {\bf A2}
(1987) 273.}

\lref\dufliib{M.~J.~Duff and J.~X.~Lu,
 Nucl. Phys. {\bf B390} (1993) 276.}

\lref\nictv{H. Nicolai, P. K. Townsend and P. van
Nieuwenhuizen, Lett. Nuovo
Cimento {\bf 30} (1981) 315.}

\lref\towspan{P. K. Townsend, in Proceedings of the 13th GIFT
Seminar
on Theoretical Physics: {\it Recent Problems in Mathematical
Physics} Salamanca, Spain, 15-27 June, 1992.}

\lref\dufm{M.~J.~Duff and R.~Minasian,
 Nucl. Phys. {\bf B436} (1995) 507.}

\lref\gropy{D. J. Gross, R. D. Pisarski and L. G. Yaffe,
Rev. Mod. Phys.
{\bf 53} (1981) 43.}

\lref\rohw{R. Rohm and E. Witten,
 Ann. Phys. {\bf 170} (1986) 454.}

\lref\banddf{T. Banks, M. Dine, H. Dijkstra and W. Fischler,
Phys. Lett. {\bf B212} (1988) 45.}

\lref\ferkp{S. Ferrara, C. Kounnas and M. Porrati, Phys. Lett.
{\bf B181} (1986) 263.}

\lref\ter{M. Terentev, Sov. J. Nucl. Phys. {\bf 49} (1989) 713.}

\lref\hass{S. F. Hassan and A. Sen, Nucl. Phys. {\bf B375}
(1992) 103.}

\lref\mahs{J. Maharana and J. Schwarz, Nucl. Phys. {\bf B390}
(1993) 3.}

\lref\senrev{A. Sen,
 Int. J. Mod. Phys. {\bf A9} (1994) 3707.}

\lref\senone{A.~Sen,
Nucl. Phys. {\bf B404} (1993) 109.}

\lref\sentwo{A.~Sen,
  Int. J. Mod. Phys. {\bf A8} (1993) 5079.}

\lref\schsen{J.~H.~Schwarz and A.~Sen,
Nucl. Phys. {\bf B411} (1994) 35.}

\lref\schsentwo{J.~H.~Schwarz and A.~Sen,
 Phys. Lett. {\bf B312} (1993) 105.}

\lref\schtwo{J.~Schwarz,
CALT-68-1815.}

\lref\senph{A.~Sen, Phys. Lett. {\bf B303} (1993) 22.}

\lref\schwarz{J.~H.~Schwarz, CALT-68-1879, hepth/9307121.}

\lref\dufldr{M. J. Duff and J. X. Lu, Nucl. Phys. {\bf B347}
(1990) 394.}

\lref\salstr{A. Salam and J. Strathdee, Phys. Lett. {\bf B61}
(1976) 375.}

\lref\jjj{J. Gauntlett, J. Harvey and J. T. Liu,
Nucl. Phys. {\bf B409} (1993) 363.}

\lref\dupo{M.~J.~Duff and C.~N.~Pope, Nucl. Phys {\bf B255}
(1985) 355.}

\lref\sg{E.~Cremmer, S.~Ferrara, L.~Girardello and
 A.~Van Proeyen,
 Phys. Lett. {\bf B116} (1982) 231.}

\lref\sgone{E.~Cremmer, S.~Ferrara, L.~Girardello and
 A.~Van Proeyen,
 Nucl. Phys. {\bf B212} (1983) 413.}

\lref\fkp{S. Ferrara, C. Kounnas and M. Porrati,
Phys. Lett. {\bf B181} (1986) 263.}

\lref\fp{S. Ferrara and M. Porrati,
Phys. Lett. {\bf B216} (1989) 289.}

\lref\ghs{D.~Garfinkle, G.~T.~Horowitz and A.~Strominger,
Phys. Rev. {\bf D43}
(1991) 3140.}

\lref\hor{G.~T.~Horowitz, in Proceedings of Trieste '92,
{\it String theory and quantum gravity '92} p.55.}

\lref\gidps{S.~B.~Giddings, J.~Polchinski and A.~Strominger,
Phys. Rev. {\bf D48} (1993) 5784.}

\lref\shatw{A.~Shapere, S.~Trivedi and F.~Wilczek,
 Mod. Phys. Lett. {\bf A6}
(1991) 2677.}

\lref\klopv{R.~Kallosh, A.~Linde, T.~Ortin, A.~Peet and
A.~Van~Proeyen,
       Phys. Rev. {\bf D46} (1992) 5278.}

\lref\kal{R.~Kallosh, Phys. Lett. {\bf B282} (1992) 80.}

\lref\ko{R.~Kallosh and T.~Ortin, Phys. Rev. {\bf D48} (1993)
742.}

\lref\hw{C.~F.~E.~Holzhey and F.~Wilczek, Nucl. Phys.
{\bf B360} (1992) 447.}

\lref\dauria{R.~D'Auria, S.~Ferrara and M.~Villasante,
Class. Quant. Grav. {\bf 11} (1994) 481.}

\lref\gibp{G.~W.~Gibbons and M.~J.~Perry, Nucl. Phys.
{\bf B248} (1984) 629.}

\lref\salam{S. W. Hawking, Monthly Notices Roy. Astron. Soc.
 {\bf 152} (1971) 75; Abdus Salam in
   {\it Quantum Gravity: an Oxford Symposium} (Eds. Isham,
Penrose
and Sciama, O.U.P. 1975); G. 't Hooft, Nucl. Phys. {\bf B335}
(1990) 138.}

\lref\susskind{ L. Susskind, RU-93-44, hepth/9309145;
   J. G. Russo and L. Susskind, UTTG-9-94,
hepth/9405117.}

\lref\gibbons{ G. W. Gibbons, in {\it Supersymmetry,
Supergravity
and Related Topics}, Eds. F. del Aguila, J. A. Azcarraga and
L. E. Ibanez
(World Scientific, 1985).}

\lref\aichelburg{ P. Aichelburg and F. Embacher, Phys. Rev.
{\bf D37}
(1986) 3006.}

\lref\geroch{ R. Geroch, J. Math. Phys. {\bf 13} (1972) 394.}

\lref\hosoya{ A. Hosoya, K.
Ishikawa, Y. Ohkuwa and K. Yamagishi, Phys. Lett. {\bf B134}
(1984) 44.}

\lref\gibw{ G. W. Gibbons
and D. L. Wiltshire, Ann. of Phys. {\bf 167} (1986) 201.}

\lref\senprl{ A. Sen, Phys. Rev. Lett. {\bf 69}
(1992) 1006.}

\lref\schild{ G. C. Debney, R. P. Kerr and
   A. Schild, J. Math. Phys. {\bf 10} (1969) 1842.}

\lref\hort{G. T. Horowitz and A. A. Tseytlin, Phys. Rev.
{\bf D50} (1994) 5204.}

\lref\hortsey{G. T. Horowitz and A. A. Tseytlin,
Imperial/TP/93-94/51, UCSBTH-94-24, hepth/9408040;
Imperial/TP/93-94/54, UCSBTH-94-31, hepth/9409021.}

\lref\cvey{M. Cveti\v c and D. Youm, UPR-623-T, hepth/9409119.}

\lref\tseytlin{A. A. Tseytlin, Imperial-TP-93-94-46,
hepth/9407099.}

\lref\klim{C. Klimcik and A. A. Tseytlin, Nucl. Phys.
{\bf B424} (1994) 71.}

\vsize=9truein
\hsize=6.5truein
\baselineskip15pt

\newsec{Introduction}

The construction of soliton solutions in string
theory is intimately connected with the presence of various
dualities in string theory (for recent reviews of string
solitons, see \refs{\bhs,\prep}).
Most of the solitonic solutions found
so far break half of the spacetime supersymmetries
of the theory in which they arise. Examples
of string-like solitons (i.e. with one Killing direction)
in this class are
the fundamental string solution of \dabghr\ and the dual
string solution of \dufkexst, which are interchanged once
the roles of the strong/weak coupling
$S$-duality and target space $T$-duality are interchanged.

In this talk I will first summarize the basic features of $S$ duality,
$T$ duality and string/string duality in heterotic string theory.
Then I will discuss
newly constructed \duffkr\ classes of string-like soliton solutions,
making connections between the solution-generating subgroup of the
$T$-duality group and the number of spacetime
supersymmetries broken in $N=4$, $D=4$ compactifications
of $N=1$, $D=10$ heterotic string theory, as well as the natural
realization of these solutions in $N=1$ and $N=2$ four-dimensional
 compactifications.
For simplicity, I will restrict myself to solutions in the gravitational
sector of the string (i.e. all Yang-Mills fields will be set to zero).

For an interesting discussion of six-dimensional string/string
duality see \dufnew.
New and exciting connections between the various
dualities in heterotic string theory and type II string theory can be
found in \hult.

\newsec{$S$ Duality}

We adopt the following conventions for
$N=1$, $D=10$ heterotic string theory compactified to
$N=4$, $D=4$
heterotic string theory: $(0123)$ is the
four-dimensional spacetime, $z=x_2+ix_3=re^{i\theta}$,
$(456789)$ are
the compactified directions, $S=e^{-2\Phi} + ia=S_1+iS_2$, where
$\Phi$ and $a$ are the four-dimensional dilaton and axion.
$S$ duality generalizes strong-weak coupling duality, since
$g=e^{\Phi}$ is the string loop coupling parameter. In $N=4$, $D=4$
heterotic string theory $S$ duality corresponds to the
group $SL(2,Z)$. In other words, the four-dimensional theory
exhibits an invariance under
\eqn\sduality{S\to {aS+b\over cS+d},}
where $a,b,c,d$ are integers and $ad-bc=1$.

There is now considerable evidence
\refs{\fonilq\reyone\senone
\senph\senrev\senmod\schsentwo{--}\schtwo,\dufkexst,\dufr}\
in favor of $S$ duality also being an exact symmetry of the full
string theory. One obvious attraction
to demonstrating $S$ duality exactly in string theory is that it would
allow
us to use perturbative string techniques in the strong-coupling regime.

In the absence
of nontrivial moduli and Yang-Mills fields, the low-energy
four-dimensional bosonic effective action in the gravitational
sector of the heterotic string has the form
\eqn\sfours{S_4=\int d^4x \sqrt{-g}\biggl(R-
{g^{\mu\nu}\over
2S^2_1}\partial_\mu S \partial_\nu \bar S  \biggr).}
A solution of this action is given by \dabghr
\eqn\sstring{\eqalign{
ds^2&=-dt^2+dx_1^2 + Re S (dx_2^2 + dx_3^2) \cr
 S& = -{1\over 2\pi} \sum_{i=1}^N n_i
\ln {(z-a_i)\over r_{i0}} ,\cr}}
where $N$ is an
arbitrary number of string-like solitons
each with arbitrary location $a_i$ in the complex
$z$-plane and
arbitrary winding number $n_i$
respectively. One can replace $z$ by $\bar z$
in $S$, thereby changing the orientation of the windings.
There is also an $SL(2,R)$ symmetry manifest in the low-energy
action, which is broken down to $SL(2,Z)$ in string theory
via axion quantization
 and from which the
above solution can be generalized further. Note that the $x_1$
Killing direction gives the above solution the structure of
a parallel multi-string configuration. Each string is
interpreted as a macroscopic fundamental string \dabghr. For
dynamical evidence for this identification see \khuscat.

\newsec{$T$ Duality}

$T$ duality in string theory is the target space duality, and
generalizes the $R\to \alpha'/R$ duality in compactified
string theory.
For $N=4$, $D=4$ compactifications of heterotic string theory,
$T$-duality corresponds to the discrete group $O(6,22;Z)$
and is known to be an
exact symmetry of the full string theory
\refs{\kikyam\sakai\bush\nair\duff\tseyv\tseycqg{--}\givpr}.

Let us consider a
simple special compactification, in which the only nontrivial
moduli are given by $T=T_1 +iT_2=\sqrt{\det g_{mn}}-i B_{45}$,
where $m,n=4,5$.
For trivial $S$ field, the low-energy
four-dimensional bosonic effective action in the gravitational
sector has the form
\eqn\sfourt{S_4=\int d^4x \sqrt{-g}\biggl(R-
{g^{\mu\nu}\over
2T^2_1}\partial_\mu T \partial_\nu \bar T \biggr).}
A solution of this action is given by \dufkexst
\eqn\sstring{\eqalign{
ds^2&=-dt^2+dx_1^2 + Re T (dx_2^2 + dx_3^2) \cr
 T& = -{1\over 2\pi} \sum_{j=1}^M m_i
\ln {(z-b_j)\over r_{j0}} ,\cr}}
where $M$ is an
arbitrary number of string-like solitons
each with arbitrary location $b_j$ in the complex
$z$-plane and
arbitrary winding number  $m_j$
respectively. Again, one can replace $z$ by $\bar z$
in $T$ and reverse the windings, and
there is an $SL(2,R)$ symmetry manifest in the low-energy
action which is broken down to $SL(2,Z)$, this time due to the
presence of instantons, and from which the
above solution can be generalized further. Note that the $x_1$
Killing direction gives the above solution the structure of
a parallel multi-string configuration as well, but in this case each
string is
interpreted as a dual string \dufkexst.

\newsec{String/String Duality}

Note that in
interchanging the $S$ field in the action \sfours\ with the $T$
field in the action \sfourt, one is interchanging the $S$ (fundamental)
string with the $T$ (dual) string and effectively interchanging
their respective couplings.
In this form, the string/string duality conjecture postulates the
existence of a dual string theory, in which the roles of the
strong/weak coupling duality and target space duality are interchanged.
It follows that the string/string duality conjecture requires the
interchange of worldsheet coupling associated with $T$ duality
and spacetime coupling associated with $S$ duality.

Of course the full $T$ duality group $O(6,22;Z)$ is much larger
than the $S$ duality group $SL(2,Z)$, but from the six-dimensional
viewpoint, the strong/weak coupling duality of the fundamental string
can be seen to emerge as a subgroup of the target space duality group
of the dual string. From the ten-dimensional viewpoint, the dual
theory is a theory of fundamental fivebranes ($5+1$-dimensional
objects) \duflrsfd. However, given
the various difficulties in working with fundamental
fivebranes (see discussion in \dufnew) and the fact that the technology
of fundamental string theory is reasonably well-developed, it seems
natural to prefer string/string duality (in $D=6$ or $D=4$) over
string/fivebrane duality in $D=10$.

Both the fundamental ($S$) and dual ($T$) string break $1/2$ the
spacetime supersymmetries, which can be seen either from the
$N=1$, $D=10$ uncompactified theory or from the $N=4$, $D=4$
compactified theory. They also both
arise in a larger $O(8,24;Z)$ solution generating group
(for an explicit $O(8,24;Z)$ transformation that takes one from
the fundamental string to the dual string see \duffkr).
As a consequence, they both saturate Bogomol'nyi bounds
and correspond to states in the
spectrum of the theory \refs{\senrev,\dufr}.

\newsec{Generalized Solutions and Supersymmetry Breaking}

Now consider the following ansatz, in which the solution-generating
subgroup of the $O(6,22;Z)$ $T$ duality group is contained in
$SL(2,Z)^3=SL(2,Z) \times SL(2,Z) \times SL(2,Z)$:
\eqn\tdefn{\eqalign{T^{(1)}&=T^{(1)}_1+iT^{(1)}_2=\sqrt{{\rm det}
g_{mn}}-iB_{45}, \quad\quad m,n=4,5,\cr
T^{(2)}&=T^{(2)}_1+iT^{(2)}_2=\sqrt{{\rm det} g_{pq}}-iB_{67},
\quad\quad p,q=6,7,\cr
T^{(3)}&=T^{(3)}_1+iT^{(3)}_2=\sqrt{{\rm det} g_{rs}}-iB_{89},
\quad\quad r,s=8,9\cr}}
are the moduli.
 We assume dependence only on the
coordinates $x_2$ and $x_3$ (i.e. $x_1$ remains a
Killing direction), and that no other moduli than the ones above
are nontrivial.

The canonical four-dimensional bosonic action for the
above compactification ansatz in the
gravitational sector can be written in terms of $g_{\mu\nu}$
($\mu,\nu=0,1,2,3$), $S$ and $T^{(a)}, a=1,2,3$ as
\eqn\sfour{\eqalign{S_4=\int d^4x \sqrt{-g}\biggl(&R-
{g^{\mu\nu}\over
2S^2_1}\partial_\mu S \partial_\nu \bar S \cr &-
{g^{\mu\nu}\over
2T^{(1)^2}_1}\partial_\mu T^{(1)}
\partial_\nu \bar T^{(1)} - {g^{\mu\nu}\over
2T^{(2)^2}_1}\partial_\mu T^{(2)}
\partial_\nu \bar T^{(2)} - {g^{\mu\nu}\over
2T^{(3)^2}_1}\partial_\mu T^{(3)}
\partial_\nu \bar T^{(3)} \biggr).\cr}}
A solution of this action is given by \duffkr
\eqn\sthreetten{\eqalign{
ds^2&=-dt^2+dx_1^2 + Re S Re T^{(1)} Re T^{(2)} Re T^{(3)}
(dx_2^2 + dx_3^2) \cr
 S& = -{1\over 2\pi} \sum_{i=1}^N n_i
\ln {(z-a_i)\over r_{i0}} ,\cr
T^{(1)}&=
-{1\over 2\pi} \sum_{j=1}^M m_j\ln {(z-b_j)\over r_{j0}} ,\cr
T^{(2)}&=
-{1\over 2\pi} \sum_{k=1}^P p_k\ln {(z-c_k)\over r_{k0}} ,\cr
T^{(3)}&=
-{1\over 2\pi} \sum_{l=1}^Q q_l\ln {(z-d_l)\over r_{l0}} ,\cr}}
where $N, M, P$ and $Q$ are
arbitrary numbers of string-like solitons in
$S, T^{(1)}, T^{(2)}$ and $T^{(3)}$ respectively
each with arbitrary location $a_i, b_j, c_k$ and $d_l$ in the complex
$z$-plane and
arbitrary winding number $n_i, m_j, p_k$ and $q_l$
respectively. One can replace $z$ by $\bar z$ independently
in $S$ and in each of the $T$ moduli, and in each of $S$ and the $T$ moduli
there is an $SL(2,Z)$ symmetry manifest in the action in each of the
moduli, and from which the
above solutions can be generalized further. Thus one has an overall
effective solution-generating group of $SL(2,Z)^4$.

It can be shown \duffkr\ that the solutions with trivial $S$ and
$1, 2$ and $3$ nontrivial $T$ fields preserve $1/2, 1/4$
and $1/8$ of the spacetime supersymmetries respectively, while
the solutions with nontrivial $S$ and $0, 1$ and $2$ nontrivial
$T$ fields preserve $1/2, 1/4$  and $1/8$ spacetime supersymmetries
respectively.
The solution with nontrivial $S$
and $3$ nontrivial $T$ fields preserves $1/8$ of the spacetime
supersymmetries for one chirality choice of $S$, and none of the
spacetime supersymmetries for the other, although the ansatz remains
a solution to the bosonic action in this latter case. In short,
the maximum number of
spacetime supersymmetries preserved in the $N=4$ theory for a
solution generating subgroup $SL(2,Z)^n$ of $O(8,24;Z)$ is
given by $(1/2)^n$ \duffkr.

\newsec{Discussion}

So what is the interpretation of these new solutions which break more
than half the supersymmetries, since they are not expected to arise
within the spectrum of Bogomol'nyi-saturated states in the $N=4$ theory?
It turns out that most of the above solutions that break
$1/2, 3/4$ or $7/8$ of the spacetime supersymmetries in
$N=4$ have analogs in $N=1$ or $N=2$ compactifications of heterotic
stirng theory
that break only $1/2$ the spacetime supersymmetries\footnote{$^*$}
{However, when at least one of the fields, either $S$
or one of the $T$ fields, has a different analyticity behaviour
from the rest, no supersymmetries are preserved in $N=1$ \duffkr.}.
Of course no solution actually preserves a higher total number of
supersymmetries in the lower supersymmetric theory ($N=1$ or $N=2$)
than in $N=4$, but the relative number of supersymmetries preserved
may be increased in truncating the $N=4$ theory to $N=1$ or $N=2$
by the removal of non-supersymmetric modes. Thus a solution that
preserves $1/8$ of the spacetime supersymmetreis in $N=4$ and $1/2$
of the spacetime supersymmetries in $N=1$ actually preserves the
same total amount of supersymmetry in both theories. The only difference
is that in the $N=4$ case one is starting with four times as many
supersymmetries, so that a greater number of those are broken than in
the $N=1$ case.

These solutions
are therefore in some sense realized naturally as stable
solitons only in the context of either $N=1$ or $N=2$ compactifications,
and
should lead to the construction of the Bogomol'nyi spectrum of these
theories. In these two case, however, the situation is complicated by
the absence of non-renormalization theorems present in the $N=4$ case
which guarantee the
absence of quantum corrections. An exception
to this scenario arises for $N=2$ compactifications with vanishing
$\beta$-function. The construction of these spectra remains a problem
for future research.

\vfil\eject
\listrefs
\bye